\documentclass[sigconf]{aamas}  

\usepackage{booktabs}

\usepackage{soul}
\usepackage{algorithm}
\usepackage{algpseudocode}
\usepackage[utf8]{inputenc}
\usepackage{siunitx}
\usepackage{amssymb}
\usepackage{amsmath}
\usepackage{listings}
\usepackage{color,soul}
\usepackage{multirow}
\usepackage{setspace}
\usepackage[framemethod=tikz]{mdframed}
\usepackage[justification=centering]{caption}
\usepackage{hyperref}
\usepackage{subcaption}
\usepackage{graphicx}
\graphicspath{ {images/} }
\usepackage{afterpage}
\usepackage{float}
\usepackage{url}
\usepackage{longtable}
\usepackage{dirtytalk}

\DeclareMathOperator*{\argmax}{argmax} 

\definecolor{Gray}{gray}{0.85}
\definecolor{red}{RGB}{255, 0, 0}
\definecolor{orange}{RGB}{255,165,0}
\definecolor{green}{RGB}{0, 128, 0}
\definecolor{blue}{RGB}{0, 0, 255}

\setcopyright{ifaamas}  
\acmDOI{doi}  
\acmISBN{}  
\acmConference[AAMAS'18]{Proc.\@ of the 17th International Conference on Autonomous Agents and Multiagent Systems (AAMAS 2018), M.~Dastani, G.~Sukthankar, E.~Andre, S.~Koenig (eds.)}{July 2018}{Stockholm, Sweden}  
\acmYear{2018}  
\copyrightyear{2018}  
\acmPrice{}  

\begin{document}

\title{Lenient Multi-Agent Deep Reinforcement Learning}  




%
\author{Gregory Palmer}
\affiliation{
  \institution{University of Liverpool}
  \country{United Kingdom}
}
\email{G.J.Palmer@liverpool.ac.uk}

\author{Karl Tuyls}
\affiliation{
  \institution{DeepMind and University of Liverpool}
  \country{United Kingdom}
}
\email{karltuyls@google.com}

\author{Daan Bloembergen}
\affiliation{
  \institution{Centrum Wiskunde \& Informatica}
  \country{The Netherlands}
}
\email{d.bloembergen@cwi.nl}

\author{Rahul Savani}
\affiliation{
  \institution{University of Liverpool}
  \country{United Kingdom}
}
\email{Rahul.Savani@liverpool.ac.uk}

%
%
%
%
%
%

\begin{abstract} 

Much of the success of single agent deep reinforcement learning (DRL) in recent years can be attributed to the use of experience replay memories (ERM), which allow Deep Q-Networks (DQNs) to be trained efficiently through sampling stored state transitions. However, care is required when using ERMs for multi-agent deep reinforcement learning (MA-DRL), as stored transitions can become outdated because agents update their policies in parallel {\cite{foerster2017stabilising}}. In this work we apply \emph{leniency} {\cite{panait2006lenient}} to MA-DRL. Lenient agents map state-action pairs to decaying temperature values that control the amount of leniency applied towards negative policy updates that are sampled from the ERM. This introduces optimism in the value-function update, and has been shown to facilitate cooperation in tabular fully-cooperative multi-agent reinforcement learning problems. We evaluate our Lenient-DQN (LDQN) empirically against the related Hysteretic-DQN (HDQN) algorithm {\cite{omidshafiei2017deep}} as well as a modified version we call \textit{scheduled}-HDQN, that uses average reward learning near terminal states. Evaluations take place in extended variations of the Coordinated Multi-Agent Object Transportation Problem (CMOTP) {\cite{bucsoniu2010multi}} which include fully-cooperative sub-tasks and stochastic rewards. We find that LDQN agents are more likely to converge to the optimal policy in a stochastic reward CMOTP compared to standard and scheduled-HDQN agents.
\end{abstract}

%


\maketitle



\section{Introduction}  \label{section:introduction}


The field of \emph{deep reinforcement learning} has seen a great number of successes in recent years. Deep reinforcement learning agents have been shown to master numerous complex problem domains, ranging from computer games \cite{mnih2015human,lample2017playing,van2016deep,schaul2015prioritized} to robotics tasks \cite{de2015importance,gu2016deep}. Much of this success can be attributed to using convolutional neural network (\emph{ConvNet}) architectures as function approximators, allowing reinforcement learning agents to be applied to domains with large or continuous state and action spaces. ConvNets are often trained to approximate policy and value functions through sampling past state transitions stored by the agent inside an \emph{experience replay memory} (ERM). 

Recently the sub-field of multi-agent deep reinforcement learning (MA-DRL) has received an increased amount of attention. Multi-agent reinforcement learning is known for being challenging even in environments with only two implicit learning agents, lacking the convergence guarantees present in most single-agent learning algorithms \cite{matignon2012independent,bloembergen2015jair}. One of the key challenges faced within multi-agent reinforcement learning is the \emph{moving target problem}: Given an environment with multiple agents whose rewards depend on each others' actions, the difficulty of finding optimal policies for each agent is increased due to the policies of the agents being non stationary \cite{busoniu2008comprehensive,TuylsW12,hernandez2017survey}. The use of an ERM amplifies this problem, as a large proportion of the state transitions stored can become deprecated \cite{omidshafiei2017deep}. 

Due to the moving target problem reinforcement learning algorithms that converge in a single agent setting often fail in fully-cooperative multi-agent systems with independent learning agents that require implicit coordination strategies. Two well researched approaches used to help parallel reinforcement learning agents overcome the moving target problem in these domains include hysteretic Q-learning \cite{matignon2007hysteretic} and leniency \cite{panait2006lenient}. Recently Omidshafiei et al. \cite{omidshafiei2017deep} successfully applied concepts from hysteretic Q-learning to MA-DRL (HDQN). However, we find that standard HDQNs struggle in fully cooperative domains that yield stochastic rewards. In the past lenient learners have been shown to outperform hysteretic agents for fully cooperative stochastic games within a tabular setting \cite{JMLR:v17:15-417}. This raises the question whether leniency can be applied to domains with a high-dimensional state space.

In this work we show how \textit{lenient learning} can be extended to MA-DRL. Lenient learners store temperature values that are associated with state-action pairs. Each time a state-action pair is visited the respective temperature value is decayed, thereby decreasing the amount of leniency that the agent applies when performing a policy update for the state-action pair. The stored temperatures enable the agents to gradually transition from optimists to average reward learners for frequently encountered state-action pairs, allowing the agents to outperform optimistic and maximum based learners in environments with misleading stochastic rewards \cite{JMLR:v17:15-417}. We extend this idea to MA-DRL by storing leniency values in the ERM, and demonstrate empirically that lenient MA-DRL agents that learn implicit coordination strategies in parallel are able to converge on the optimal joint policy in difficult coordination tasks with stochastic rewards. We also demonstrate that the performance of Hysteretic Q-Networks (HDQNs) within stochastic reward environments can be improved with a scheduled approach.

Our main contributions can be summarized as follows. \\
\textbf{1)} We introduce the Lenient Deep Q-Network (LDQN) algorithm which includes two extensions to leniency: a retroactive \emph{temperature decay schedule} (TDS) that prevents premature temperature cooling, and a $\overline{T}(s)$-Greedy exploration strategy, where the probability of the optimal action being selected is based on the average temperature of the current state. When combined, TDS and $\overline{T}(s)$-Greedy exploration encourage exploration until average rewards have been established for later transitions.\\
\textbf{2)} We show the benefits of using TDS over \emph{average temperature folding} (ATF) {\cite{JMLR:v17:15-417}}. \\
\textbf{3)} We provide an extensive analysis of the leniency-related hyperparameters of the LDQN. \\
\textbf{4)} We propose a \textit{scheduled}-HDQN that applies less optimism towards state transitions near terminal states compared to earlier transitions within the episode. \\
\textbf{5)} We introduce two extensions to the Cooperative Multi-agent Object Transportation Problem (CMOTP) {\cite{bucsoniu2010multi}}, including narrow passages that test the agents' ability to master fully-cooperative sub-tasks, stochastic rewards and noisy observations. \\
\textbf{6)} We empirically evaluate our proposed LDQN and SHDQN against standard HDQNs using the extended versions of the CMOTP. We find that while HDQNs perform well in deterministic CMOTPs, they are significantly outperformed by SHDQNs in domains that yield a stochastic reward. Meanwhile LDQNs comprehensively outperform both approaches within the stochastic reward CMOTP. 

The paper proceeds as follows: first we discuss related work and the motivation for introducing leniency to MA-DRL. We then provide the reader with the necessary background regarding how leniency is used within a tabular setting, briefly introduce hysteretic Q-learning and discuss approaches for clustering states based on raw pixel values. We subsequently introduce our contributions, including the Lenient-DQN architecture, $\overline{T}$(s)-Greedy exploration, Temperature Decay Schedules, Scheduled-HDQN and extensions to the CMOTP, before moving on to discuss the results of empirically evaluating the LDQN. Finally we summarize our findings, and discuss future directions for our research.


\section{Related work}

A number of methods have been proposed to help deep reinforcement learning agents converge towards an optimal joint policy in cooperative multi-agent tasks. Gupta et al. \cite{guptacooperative} evaluated policy gradient, temporal difference error, and actor critic methods on cooperative control tasks that included discrete and continuous state and action spaces, using a decentralized parameter sharing approach with centralized learning. In contrast our current work focuses on decentralized-concurrent learning. A recent successful approach has been to decompose a team value function into agent-wise value functions through the use of a value decomposition network architecture \cite{sunehag2017value}. Others have attempted to help concurrent learners converge through identifying and deleting obsolete state transitions stored in the replay memory. For instance, Foerster et al. \cite{foerster2017stabilising} used importance sampling as a means to identify outdated transitions while maintaining an action observation history of the other agents. Our current work does not require the agents to maintain an action observation history. Instead we focus on optimistic agents within environments that require implicit coordination. This decentralized approach to multi-agent systems offers a number of advantages including speed, scalability and robustness \cite{matignon2012independent}. The motivation for using implicit coordination is that communication can be expensive in practical applications, and requires efficient protocols \cite{matignon2012independent, tan1993multi, balch1994communication}. 

Hysteretic Q-learning is a form of optimistic learning with a strong empirical track record in fully-observable multi-agent reinforcement learning \cite{matignon2012independent, barbalios2014robust, xu2012multiagent}. Originally introduced to prevent the overestimation of Q-Values in stochastic games, hysteretic learners use two learning rates: a learning rate $\alpha$ for updates that increase the value estimate (Q-value) for a state-action pair and a smaller learning rate $\beta$ for updates that decrease the Q-value \cite{matignon2007hysteretic}. However, while experiments have shown that hysteretic learners perform well in deterministic environments, they tend to perform sub-optimally in games with stochastic rewards. Hysteretic learners' struggles in these domains have been attributed to learning rate $\beta$'s inter-dependencies with the other agents' exploration strategies \cite{matignon2012independent}. 

Lenient learners present an alternative to the hysteretic approach, and have empirically been shown to converge towards superior policies in stochastic games with a small state space \cite{JMLR:v17:15-417}. Similar to the hysteretic approach, lenient agents initially adopt an optimistic disposition, before gradually transforming into average reward learners \cite{JMLR:v17:15-417}. 
Lenient methods have received criticism in the past for the time they require to converge {\cite{JMLR:v17:15-417}}, the difficulty involved in selecting the correct hyperparameters, the additional overhead required for storing the temperature values, and the fact that they were originally only proposed for matrix games \cite{matignon2012independent}. However, given their success in tabular settings we here investigate whether leniency can be applied successfully to MA-DRL.

\section{Background} \label{section:background}

\paragraph{\bf Q-Learning.} \label{section:background:QLearning}

The algorithms implemented for this study are based upon Q-learning, a form of temporal difference reinforcement learning that is well suited for solving sequential decision making problems that yield stochastic and delayed rewards \cite{watkins1992q,sutton1998reinforcement}. The algorithm learns Q-values for state-action pairs which are estimates of the discounted sum of future rewards (the return) that can be obtained at time $t$ through selecting action $a_t$ in a state $s_t$, providing the optimal policy is selected in each state that follows.

Since most interesting sequential decision problems have a large state-action space, Q-values are often approximated using function approximators such as tile coding \cite{sutton1998reinforcement} or neural networks \cite{van2016deep}. The parameters $\theta$ of the function approximator can be learned from experience gathered by the agent while exploring their environment, choosing an action $a_{t}$ in state $s_{t}$ according to a policy $\pi$, and updating the Q-function by bootstrapping the immediate reward $r_{t+1}$ received in state $s_{t+1}$ plus the expected future reward from the next state (as given by the Q-function):
\begin{equation} \label{eq:parameterUpdate}
\theta_{t+1} = \theta_{t} + \alpha \big(Y^{Q}_{t} - Q(s_{t}, a_{t};\theta_t)\big) \nabla_{\theta_t}Q(s_t, a_t; \theta_t).
\end{equation}
Here, $Y_{t}^{Q}$ is the bootstrap target which sums the immediate reward $r_{t+1}$ and the current estimate of the return obtainable from the next state $s_{t+1}$ assuming optimal behaviour (hence the $\max$ operator) and discounted by $\gamma \in (0, 1]$, given in Eq. \eqref{eq:targetValue}. The Q-value $Q(s_{t}, a_{t}; \theta_t)$ moves towards this target by following the gradient $\nabla_{\theta_t}Q(s_t, a_t; \theta_t)$; $\alpha \in (0,1]$ is a scalar used to control the learning rate.
\begin{equation} \label{eq:targetValue}
Y^{Q}_{t} \equiv r_{t+1} + \gamma \max_{a \in A} Q(s_{t+1}, a; \theta_t).
\end{equation}

\paragraph{\bf Deep Q-Networks (DQN)}

In deep reinforcement learning \cite{mnih2015human} a multi-layer neural network is used as a function approximator, mapping a set of $n$-dimensional state variables to a set of $m$-dimensional Q-values $f:\mathbb{R}^n \to \mathbb{R}^m$, where m represents the number of actions available to the agent. The network parameters $\theta$ can be trained using stochastic gradient descent, randomly sampling past transitions experienced by the agent that are stored within an experience replay memory (ERM) \cite{lin1992self,mnih2015human}. Transitions are tuples $(s_t,a_t,s_{t+1},r_{t+1})$ consisting of the original state $s_t$, the action $a_t$, the resulting state $s_{t+1}$ and the immediate reward $r_{t+1}$. The network is trained to minimize the time dependent loss function $L_i(\theta_i)$,
\begin{equation}\label{eq:time_dependent_loss_function}
L_i(\theta_i) = \mathbf{E}_{s,a \sim p(\cdot)}\Big[ (Y_t - Q(s,a; \theta_t))^2 \Big],
\end{equation}
where $p(s,a)$ represents a probability distribution of the transitions stored within the ERM, and $Y_t$ is the target:
\begin{equation} \label{eq:DQNTarget}
Y_t \equiv r_{t+1} + \gamma Q(s_{t+1}, \argmax_{a \in A} Q(s_{t+1}, a; \theta_t);\theta_{t}').
\end{equation}

Equation \eqref{eq:DQNTarget} is a form of double Q-learning \cite{hasselt2010double} in which the target action is selected using weights $\theta$, while the target value is computed using weights $\theta'$ from a target network. The target network is a more stable version of the current network, with the weights being copied from current to target network after every $n$ transitions \cite{van2016deep}. 
Double-DQNs have been shown to reduce overoptimistic value estimates \cite{van2016deep}. This notion is interesting for our current work, since both leniency and hysteretic Q-learning attempt to induce sufficient optimism in the early learning phases to allow the learning agents to converge towards an optimal joint policy. 

\paragraph{\bf Hysteretic Q-Learning.}  \label{section:background:HystereticQL}

Hysteretic Q-learning \cite{matignon2007hysteretic} is an algorithm designed for decentralised learning in deterministic multi-agent environments, and which has recently been applied to MA-DRL as well \cite{omidshafiei2017deep}. Two learning rates are used, $\alpha$ and $\beta$, with $\beta < \alpha$. The smaller learning rate $\beta$ is used whenever an update would reduce a Q-value. This results in an optimistic update function which puts more weight on positive experiences, which is shown to be beneficial in cooperative multi-agent settings. Given a spectrum with traditional Q-learning at one end and maximum-based learning, where negative experiences are completely ignored, at the other, then hysteretic Q-learning lies somewhere in between depending on the value chosen for $\beta$.

\paragraph{\bf Leniency.} \label{section:background:leniency}

Lenient learning was originally introduced by Potter and De Jong \cite{potter1994cooperative} to help cooperative co-evolutionary algorithms converge towards an optimal policy, and has later been applied to multi-agent learning as well \cite{panait2008theoretical}. It was designed to prevent \emph{relative overgeneralization} \cite{wiegand2003analysis}, which occurs when agents gravitate towards a robust but sub-optimal joint policy due to noise induced by the mutual influence of each agent's exploration strategy on others' learning updates. 

Leniency has been shown to increase the likelihood of convergence towards the globally optimal solution in stateless coordination games for reinforcement learning agents \cite{panait2006lenient,panait2008theoretical,bloembergen2011empirical}. 
Lenient learners do so by effectively forgiving (ignoring) sub-optimal actions by teammates that lead to low rewards during the initial exploration phase \cite{panait2006lenient,panait2008theoretical}. While initially adopting an optimistic disposition, the amount of leniency displayed is typically decayed each time a state-action pair is visited. As a result the agents become less lenient over time for frequently visited state-action pairs while remaining optimistic within unexplored areas. This transition to average reward learners helps lenient agents avoid sub-optimal joint policies in environments that yield stochastic rewards \cite{JMLR:v17:15-417}.

During training the frequency with which lenient reinforcement learning agents perform updates that result in lowering the Q-value of a state action pair $(s, a)$ is determined by leniency and temperature functions, $l(s_t, a_t)$ and $T_t(s_t, a_t)$ respectively \cite{JMLR:v17:15-417}. The relation of the temperature function is one to one, with each state-action pair being assigned a temperature value that is initially set to a defined maximum temperature value, before being decayed each time the pair is visited. The leniency function 
\begin{equation} \label{eq:leniency}
l(s_t, a_t) = 1 - e^{-K*T_t(s_t,a_t)}
\end{equation}
uses a constant $K$ as a leniency moderation factor to determine how the temperature value affects the drop-off in lenience. Following the update, $T_t(s_t, a_t)$ is decayed using a discount factor $\beta \in [0, 1]$ such that $T_{t+1}(s_t, a_t) = \beta\,T_t(s_t, a_t)$. 


Given a TD-Error $\delta$, where $\delta = Y_t - Q(s_t,a_t; \theta_t)$, leniency is applied to a Q-value update as follows:
\begin{equation} \label{eq:QValueUpdate}
  Q(s_t, a_t) = 
  \begin{cases}
    Q(s_t, a_t) + \alpha \delta &\text{if $\delta > 0$ or $x > l(s_t, a_t)$}.\\
    Q(s_t, a_t) &\text{if $\delta \leq 0$ and $x \leq l(s_t, a_t)$}.
  \end{cases}
\end{equation}
The random variable $x \sim U(0, 1)$ is used to ensure that an update on a negative $\delta$ is executed with a probability $1 - l(s_t, a_t)$. 

\paragraph{\bf Temperature-based exploration} \label{section:background:temperature_based_exploratoin} The temperature values maintained by lenient learners can also be used to influence the action selection policy. Recently Wei and Luke \cite{JMLR:v17:15-417} introduced Lenient Multiagent Reinforcement Learning 2 (LMRL2), where the average temperature of the agent's current state is used with the Boltzmann action selection strategy to determine the weight of each action. As a result agents are more likely to choose a greedy action within frequently visited states while remaining exploratory for less-frequented areas of the environment. However, Wei and Luke \cite{JMLR:v17:15-417} note that the choice of temperature moderation factor for the Boltzmann selection method is a non-trivial task, as Boltzmann selection is known to struggle to distinguish between Q-Values that are close together \cite{JMLR:v17:15-417,kaelbling1996reinforcement}. 

\paragraph{\bf Average Temperature Folding (ATF)} \label{section:background:avg_temperature_folding}
If the agents find themselves in the same initial state at the beginning of each episode, then after repeated interactions the temperature values for state-action pairs close to the initial state can decay rapidly as they are visited more frequently. However, it is crucial for the success of the lenient learners that the temperatures for these state-action pairs remains sufficiently high for the rewards to propagate back from later stages, and to prevent the agents from converging upon a sub-optimal policy \cite{JMLR:v17:15-417}. One solution to this problem is to fold the average temperature for the $n$ actions available to the agent in $s_{t+1}$ into the temperature that is being decayed for $(s_t, a_t)$ \cite{JMLR:v17:15-417}. 
%
%
The extent to which this average temperature $\overline{T}_t(s_{t+1}) = 1/n \sum_{i = 1}^n T_t(s_{t+1}, a_i)$ is folded in is determined by a constant $\upsilon$ as follows: 
\begin{equation}\label{temperature_update_with_diffusion}
    \hspace{-0.5em}T_{t+1}(s_t, a_t)=\beta\begin{cases}
    T_t(s_t, a_t) \hfill \text{if $s_{t+1}$ is terminal}.\\
    (1-\upsilon)T_t(s_t, a_t) + \upsilon \overline{T}_t(s_{t+1}) \hspace{1.2em} \text{otherwise}.
  \end{cases}
\end{equation}

\paragraph{\bf Clustering states using autoencoders.} \label{sec:background:clustering}

In environments with a high dimensional or continuous state space, a tabular approach for mapping each possible state-action pair to a temperature as discussed above is no longer feasible. Binning can be used to discretize low dimensional continuous state-spaces, however further considerations are required regarding mapping semantically similar states to a decaying temperature value when dealing with high dimensional domains, such as image observations. Recently, researchers studying the application of count based exploration to Deep RL have developed interesting solutions to this problem. 
For example, Tang et al. \cite{tang2016exploration} used autoencoders to automatically cluster states in a meaningful way in challenging benchmark domains including Montezuma's Revenge. 

The autoencoder, consisting of convolutional, dense, and transposed convolutional layers, can be trained using the states stored in the agent's replay memory \cite{tang2016exploration}. It then serves as a pre-processing function $g : S \rightarrow \mathbb{R}^D$, with a dense layer consisting of $D$ neurons with a saturating activation function (e.g. a Sigmoid function) at the centre. SimHash \cite{charikar2002similarity}, a locality-sensitive hashing (LSH) function, can be applied to the rounded output of the dense layer to generate a hash-key $\phi$ for a state $s$. This hash-key is computed using a constant $k \times D$ matrix $A$ with i.i.d. entries drawn from a standard Gaussian distribution $N(0,1)$ as
\begin{equation}
    \phi(s) = sgn\big(A\,g(s)\big) \in \{-1, 1\}^k.
\end{equation}
where $g(s)$ is the autoencoder pre-processing function, and $k$ controls the granularity such that higher values yield a more fine-grained clustering \cite{tang2016exploration}.

\section{Algorithmic Contributions}

In the following we describe our main algorithmic contributions. First we detail our newly proposed \textit{Lenient Deep Q-Network}, and thereafter we discuss our extension to Hysteretic DQN, which we call \textit{Scheduled HDQN}.

\subsection{Lenient Deep Q-Network (LDQN)}

\paragraph{\bf Approach} \label{section:LDQN} 

Combining leniency with DQNs requires careful considerations regarding the use of the temperature values, in particular when to compute the amount of leniency that should be applied to a state transition that is sampled from the replay memory. In our initial trials we used leniency as a mechanism to determine which transitions should be allowed to enter the ERM. However, this approach led to poor results, presumably due to the agents developing a bias during the initial random exploration phase where transitions were stored indiscriminately. To prevent this bias we use an alternative approach where we compute and store the amount of leniency at time $t$ within the ERM tuple: $(s_{t-1}, a_{t-1}, r_t, s_{t}, l(s_t, a_t)_t)$. The amount of leniency that is stored is determined by the current temperature value $T$ associated with the hash-key $\phi(s)$ for state $s$ and the selected action $a$, similar to Eq. \eqref{eq:leniency}: 
\begin{equation}
    l(s,t) = 1 - e^{-k\times T(\phi(s), a)}.
\end{equation}
We use a dictionary to map each $(\phi(s), a)$ pair encountered to a temperature value, where the hash-keys are computed using Tang et al.'s \cite{tang2016exploration} approach described in Section \ref{sec:background:clustering}. If a temperature value does not yet exist for $(\phi(s), a)$ within the dictionary then an entry is created, setting the temperature value equal to $MaxTemperature$. Otherwise the current temperature value is used and subsequently decayed, to ensure the agent will be less lenient when encountering a semantically similar state in the future. As in standard DQN the aim is to minimize the loss function of Eq. \eqref{eq:time_dependent_loss_function}, with the modification that for each sample $j$ chosen from the replay memory for which the leniency conditions of Eq. (6) are not met, are ignored.


\paragraph{\bf Retroactive Temperature Decay Schedule (TDS)} \label{section:LDQN:TDS} 

During initial trials we found that temperatures decay rapidly for state-action pairs belonging to challenging sub-tasks in the environment, even when using ATF (Section \ref{section:background:avg_temperature_folding}). In order to prevent this premature cooling of temperatures we developed an alternative approach using a pre-computed temperature decay schedule $\beta_0,\ldots,\beta_{n}$ with a step limit $n$. The values for $\beta$ are computed using an exponent $\rho$ which is decayed using a decay rate $d$: 
\begin{equation}
\beta_n = e^{\rho \times d^t}
\end{equation}
for each $t, 0 \leq t < n$.


Upon reaching a terminal state the temperature decay schedule is applied as outlined in Algorithm \ref{alg:application_of_temp_decay_schedule}. The aim is to ensure that temperature values of state-action pairs encountered during the early phase of an episode are decayed at a slower rate than those close to the terminal state transition (line 4). We find that maintaining a slow-decaying maximum temperature $\nu$ (lines 5-7) that is decayed using a decay rate $\mu$ helps stabilize the learning process when $\epsilon$-Greedy exploration is used. Without the decaying maximum temperature the disparity between the low temperatures in well explored areas and the high temperatures in relatively unexplored areas has a destabilizing effect during the later stages of the learning process. Furthermore, for agents also using the temperature values to guide their exploration strategy (see below), $\nu$ can help ensure that the agents transition from exploring to exploiting within reasonable time. The decaying maximum temperature $\nu$ is used whenever $T(\phi(s_{t-1}),a_{t-1}) > \nu_t$, or when agents fail at their task in environments where a clear distinction can be made between success and failure. Therefore TDS is best suited for domains that yield sparse large rewards.

\begin{algorithm}[tb]
\caption{Application of temperature decay schedule (TDS)}
\label{alg:application_of_temp_decay_schedule}
  \begin{algorithmic}[1]
  \State Upon reaching a terminal state \algorithmicdo
    \State $n \gets 0$, $steps \gets $ steps taken during the episode
    \For{$i = steps$ to $0$}
        \If{$\beta_n T_t(\phi(s_i), a_{i}) < \nu_t$}
        \State $T_{t+1}(\phi(s_{i}),a_{i}) \gets \beta_n T_t(\phi(s_i), a_{i})$ 
        \Else
        \State $T_{t+1}(\phi(s_{i}),a_{i}) \gets \nu_t$ 
        \EndIf
        \State $n \gets n + 1$
    \EndFor
    \State $\nu \gets \mu \, \nu$
  \end{algorithmic}
\end{algorithm}


Applying the TDS after the agents fail at a task could result in the repeated decay of temperature values for state-action pairs leading up to a sub-task. For instance, the sub-task of transporting a heavy item of goods through a doorway may only require a couple of steps for trained agents who have learned to coordinate. However, untrained agents may require thousands of steps to complete the task. If a time-limit is imposed for the agents to deliver the goods, and the episode ends prematurely while an attempt is made to solve the sub-task, then the application of the TDS will result in the rapid decay of the temperature values associated with the frequently encountered state-action pairs. We resolve this problem by setting the temperature values $T_t(\phi(s_i), a_{i}) > \nu$ to $\nu$ at the end of incomplete runs instead of repeatedly decaying them, thereby ensuring that the agents maintain a lenient disposition towards one another.



\paragraph{\bf \boldmath$\overline{T}(s_t)$-Greedy Exploration} \label{section:LDQN:T_Greedy} 

\begin{figure}[tb]
\centering
\includegraphics[width=0.4\textwidth]{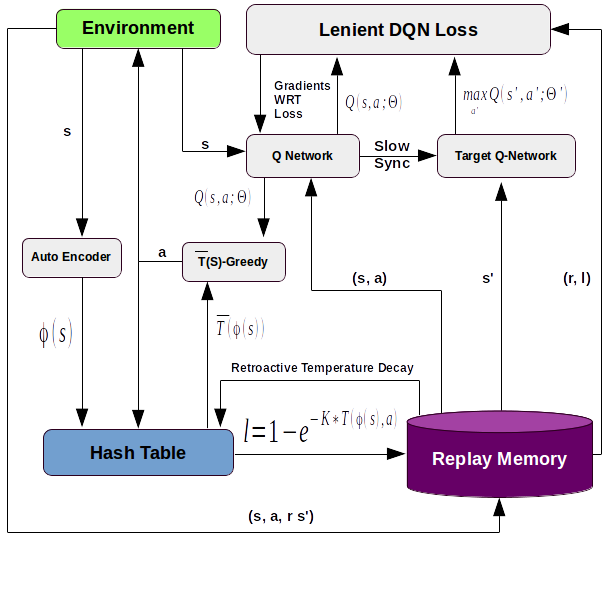} 
\captionsetup{justification=justified}
\caption{Lenient-DQN Architecture. We build on the standard Double-DQN architecture \cite{van2016deep} by adding a lenient loss function (top right, see Section \ref{section:LDQN}). Leniency values are stored in the replay memory along with the state transitions; we cluster semantically similar states using an autoencoder and SimHash (bottom left), and apply our retroactive temperature decay schedule (TDS, Algorithm \ref{alg:application_of_temp_decay_schedule}). Actions are selected using the $\overline{T}(s_t)$-Greedy exploration method.}
\label{fig:LDQN:LDQN_Diagram}
\end{figure}

During initial trials we encountered the same problems discussed by Wei and Luke \cite{JMLR:v17:15-417} regarding the selection of the temperature moderation factor for the Boltzmann action selection strategy. This led to the development of a more intuitive $\overline{T}(s_t)$-Greedy exploration method where the average temperature value $\overline{T}(s_t) \in (0, 1]$ for a state $s_t$ replaces the $\epsilon$ in the $\epsilon$-Greedy action selection method. An exponent $\xi$ is used to control the pace at which the agents transition from explorers to exploiters. The agent therefore selects action $a=argmax_aQ(a)$ with a probability $1 - \overline{T}(s_t)^\xi$ and a random action with probability $\overline{T}(s_t)^\xi$. We outline our complete LDQN architecture in Figure {\ref{fig:LDQN:LDQN_Diagram}}. 

\subsection{Scheduled-HDQN (SHDQ)}


Hysteretic Q-learners are known to converge towards sub-optimal joint policies in environments that yield stochastic rewards {\cite{JMLR:v17:15-417}}. However, drawing parallels to lenient learning, where it is desirable to decay state-action pairs encountered at the beginning of an episode at a slower rate compared to those close to a terminal state, we consider that the same principle can be applied to Hysteretic Q-learning. Subsequently we implemented \textit{Scheduled-HDQN} with a pre-computed learning rate schedule $\beta_{0}, \ldots, \beta_{n}$ where $\beta_{n}$ is set to a value approaching $\alpha$, and for each $\beta_t, 0 \leq t < n$, we have $\beta_t = d^{n - t} \beta_{n}$  using a decay coefficient $d \in (0, 1]$. The state transitions encountered throughout each episode are initially stored within a queue data-structure. Upon reaching a terminal state the $n$ state-transitions are transferred to the ERM as ($s_{t}$, $s_{t+1}$, $r_{t+1}$, $a_{t}$, $\beta_{t}$) for $t \in \{0,\ldots, n\}$. Our hypothesis is that storing $\beta$ values that approach $\alpha$ for state-transitions leading to the terminal state will help agents converge towards optimal joint policies in environments that yield sparse stochastic rewards.



\section{Empirical Evaluation} \label{section:evaluation}

\paragraph{\bf CMOTP Extensions} \label{section:evaluation:cmotp}

\begin{figure}[ht]
\centering
    \begin{subfigure}[b]{0.15\textwidth}
        \centering
        \includegraphics[width=\textwidth]{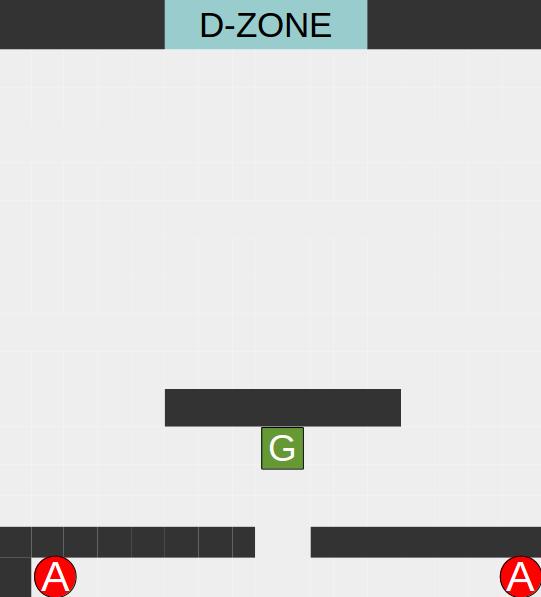} 
        \caption{Original}
        \label{fig:cmotop:original}
    \end{subfigure} 
    \begin{subfigure}[b]{0.15\textwidth}
        \centering
        \includegraphics[width=\textwidth]{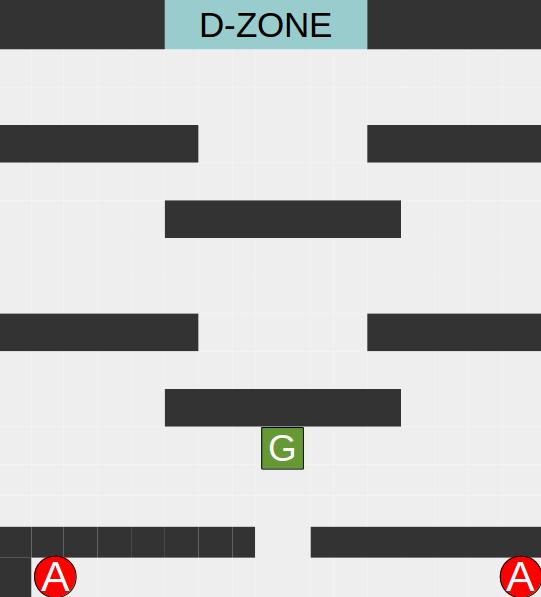} 
        \caption{Narrow-Passage}        
        \label{fig:cmotop:NarrowPassage}  
    \end{subfigure}
    \begin{subfigure}[b]{0.15\textwidth}
        \centering
        \includegraphics[width=\textwidth]{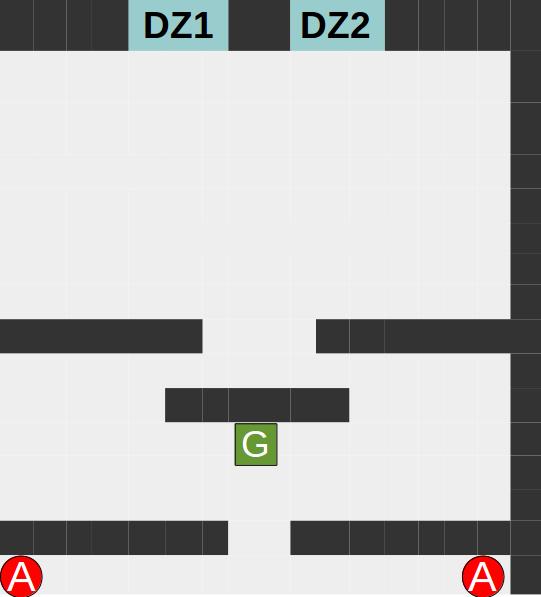} 
        \caption{Stochastic}        
        \label{fig:cmotop:StochasticReward}      
    \end{subfigure}    
\caption{CMOTP Layouts}
\label{fig:cmotp_versions}
\end{figure}

We subjected our agents to a range of Coordinated Multi-Agent Object Transportation Problems (CMOTPs) inspired by the scenario discussed in Bu{\c{s}}oniu et al. \cite{bucsoniu2010multi}, in which two agents are tasked with delivering one item of goods to a drop-zone within a grid-world. The agents must first exit a room one by one before locating and picking up the goods by standing in the grid cells on the left and right hand side. The task is fully cooperative, meaning the goods can only be transported upon both agents grasping the item and choosing to move in the same direction. Both agents receive a positive reward after placing the goods inside the drop-zone. The actions available to each agent are to either stay in place or move left, right, up or down. We subjected our agents to three variations of the CMOTP, depicted in Figure \ref{fig:cmotp_versions}, where each \emph{A} represents one of the agents, \emph{G} the goods, and \emph{D-ZONE} / \emph{DZ} mark the drop-zone(s). The layout in sub-figure \ref{fig:cmotop:original} is a larger version of the original layout \cite{bucsoniu2010multi}, while the layout in sub-figure \ref{fig:cmotop:NarrowPassage} introduces narrow-passages between the goods and the drop-zone, testing whether the agents can learn to coordinate in order to overcome challenging areas within the environment. The layout in sub-figure \ref{fig:cmotop:StochasticReward} tests the agents' response to stochastic rewards. Drop-zone 1 (DZ1) yields a reward of 0.8, whereas drop-zone 2 (DZ2) returns a reward of 1 on 60\% of occasions and only 0.4 on the other 40\%. DZ1 therefore returns a higher reward on average, 0.8 compared to the 0.76 returned by DZ2. A slippery surface can be added to introduce stochastic state transitions to the CMOTP, a common practice within grid-world domains where the agents move in an unintended direction with a predefined probability at each time-step. 

\paragraph{\bf Setup} \label{section:evaluation:Experiments}

We conduct evaluations using a Double-DQN architecture \cite{van2016deep} as basis for the algorithms. The Q-network consists of 2 convolutional layers with 32 and 64 kernels respectively, a fully connected layer with 1024 neurons and an output neuron for each action. The agents are fed a $16 \times 16$ tensor representing a gray-scale version of the grid-world as input. We use the following pixel values to represent the entities in our grid-world: $Agent 1 = 250$, $Agent 2 = 200$, $Goods = 150$ and $Obstacles = 50$. \emph{Adam} {\cite{kingma2014adam}} is used to optimize the networks. Our initial experiments are conducted within a noise free environment, enabling us to speed up the testing of our LDQN algorithm without having to use an autoencoder for hashing; instead we apply python's xxhash. We subsequently test the LDQN with the autoencoder for hashing in a noisy version of the stochastic reward CMOTP. The autoencoder consists of 2 convolutional Layers with 32 and 64 kernels respectively, 3 fully connected layers with 1024, 512, and 1024 neurons followed by 2 transposed convolutional layers. For our Scheduled-HDQN agents we pre-compute $\beta_{0\text{ to }n}$ by setting $\beta_n=0.9$ and applying a decay coefficient of $d=0.99$ at each step ${t = 1\text{ to } n}$, i.e. $\beta_{n-t} = 0.99^{t} \beta_n$, with $\beta_{n-t}$ being bounded below at 0.4. We summarize the remaining hyper-parameters in Table {\ref{fig:hyperparameters}}. In Section {\ref{sec:stochastic_reward_cmotp}} we include an extensive analysis of tuning the leniency related hyper-parameters. We note at this point that each algorithm used the same learning rate $\alpha$ specified in Table {\ref{fig:hyperparameters}}. 


\begin{table}[ht]
\begin{center}
{\scriptsize
\begin{tabular}{ | c | c | c |}
\hline
\textbf{Component} & \textbf{Hyper-parameter} & \textbf{Setting} \\
\hline
\hline
\multirow{4}{*}{\textbf{DQN-Optimization}} & Learning rate $\alpha$ & 0.0001 \\
\cline{2-3}   
& Discount rate $\gamma$ & 0.95 \\
\cline{2-3}   
& Steps between target network synchronization & 5000  \\
\cline{2-3}   
& ERM Size &  250'000 \\
\hline
\hline
\multirow{ 3}{*}{\textbf{$\epsilon$-Greedy Exploration}} & Initial $\epsilon$ value & 1.0 \\
\cline{2-3}   
& $\epsilon$ Decay factor & 0.999 \\
\cline{2-3}   
& Minimum $\epsilon$ Value & 0.05\\
\hline
\hline
\multirow{6}{*}{\textbf{Leniency}} & MaxTemperature & 1.0 \\
\cline{2-3}   
& Temperature Modification Coefficient $K$ & 2.0 \\
\cline{2-3}   
& TDS Exponent $\rho$ & -0.01 \\
\cline{2-3}   
& TDS Exponent Decay Rate $d$ & 0.95\\
\cline{2-3}   
& Initial Max Temperature Value $\nu$ & 1.0 \\
\cline{2-3}   
& Max Temperature Decay Coefficient $\mu$ & 0.999 \\
\hline  
\hline
\multirow{2}{*}{\textbf{Autoencoder}} & HashKey Dimensions $k$ & 64 \\
\cline{2-3}   
& Number of sigmoidal units in the dense layer $D$ & 512 \\
\hline  
\end{tabular}}
\end{center}
\captionsetup{justification=centering}
\caption{Hyper-parameters}
\label{fig:hyperparameters}
\end{table}   

\section{Deterministic CMOTP Results} \label{Deterministic_CMOTP_Results}

\paragraph{\bf Original CMOTP} 

The CMOTP represents a challenging fully cooperative task for parallel learners. Past research has shown that deep reinforcement learning agents can converge towards cooperative policies in domains where the agents receive feedback for their individual actions, such as when learning to play pong with the goal of keeping the ball in play for as long as possible {\cite{tampuu2017multiagent}}. However, in the CMOTP feedback is only received upon delivering the goods after a long series of coordinated actions. No immediate feedback is available upon miscoordination. When using uniform action selection the agents only have a 20\% chance of choosing identical actions per state transition. As a result thousands of state transitions are often required to deliver the goods and receive a reward while the agents explore the environment, preventing the use of a small replay memory where outdated transitions would be overwritten within reasonable time. As a result standard Double-DQN architectures struggled to master the CMOTP, failing to coordinate on a significant number of runs even when confronted with the relatively simple original CMOTP.

We conducted 30 training runs of 5000 episodes per run for each LDQN and HDQN configuration. Lenient and hysteretic agents with $\beta < 0.8$ fared significantly better than the standard Double-DQN, converging towards joint policies that were only a few steps shy of the optimal 33 steps required to solve the task. Lenient agents implemented with both ATF and TDS delivered a comparable performance to the hysteretic agents with regards to the average steps per episode and the coordinated steps percentage measured over the final 100 steps of each episode (Table \ref{fig:originalCMOTP:Results}, left). However, both LDQN-ATF and LDQN-TDS averaged a statistically significant higher number of steps per training run compared to hysteretic agents with $\beta < 0.7$. For the hysteretic agents we observe a statistically significant increase in the average steps per run as the values for $\beta$ increase, while the average steps and coordinated steps percentage over the final 100 episodes remain comparable. 



\begin{table*}[t!]
\begin{center}
{\scriptsize
\begin{tabular}{ | l | c | c | c | c | c | c | c | c | c | c |}
\cline{2-11}   
\multicolumn{1}{c |}{} & \multicolumn{6}{c|}{\textbf{Original CMOTP Results}} & \multicolumn{4}{ c|}{\textbf{Narrow-Passage CMOTP Results}}  \\
\cline{2-11}   
\multicolumn{1}{c |}{}  & \textbf{Hyst. \boldmath$\beta=0.5$} & \textbf{Hyst. \boldmath$\beta=0.6$} & \textbf{Hyst. \boldmath$\beta=0.7$} & \textbf{Hyst. \boldmath$\beta=0.8$} & \textbf{LDQN ATF}  & \textbf{LDQN TDS}  & \textbf{Hyst. \boldmath$\beta=0.5$} & \textbf{Hyst. \boldmath$\beta=0.6$} & \textbf{LDQN ATF}  & \textbf{LDQN TDS}  \\
 \cline{2-10}   
\hline  
\textbf{SPE}  & 36.4 & 36.1 & 36.8  & 528.9 &36.9 & 36.8 & 45.25 & 704.9 & 376.2 & 45.7 \\
\hline  
\textbf{CSP}& 92\% & 92\% & 92\% & 91\% & 92\% & 92\% &  92\% & 89\%  & 90\% & 92\% \\
\hline 
\textbf{SPR} & 1'085'982 & 1'148'652 & 1'408'690 & 3'495'657 & 1'409'720 & 1'364'029 & 1'594'968 & 4'736'936 & 3'950'670  & 2'104'637 \\
\hline 
\end{tabular}}
\end{center}
\captionsetup{justification=centering}
\caption{Deterministic CMTOP Results, including average steps per episode (\textbf{SPE}) over the final 100 episodes, coordinated steps percentages (\textbf{CSP}) over the final 100 episodes, and the average steps per training run (SPR). }
\label{fig:originalCMOTP:Results}
\end{table*}  

\paragraph{\bf Narrow Passage CMOTP}

\begin{figure}[bh]
\centering
\includegraphics[width=0.35\textwidth]{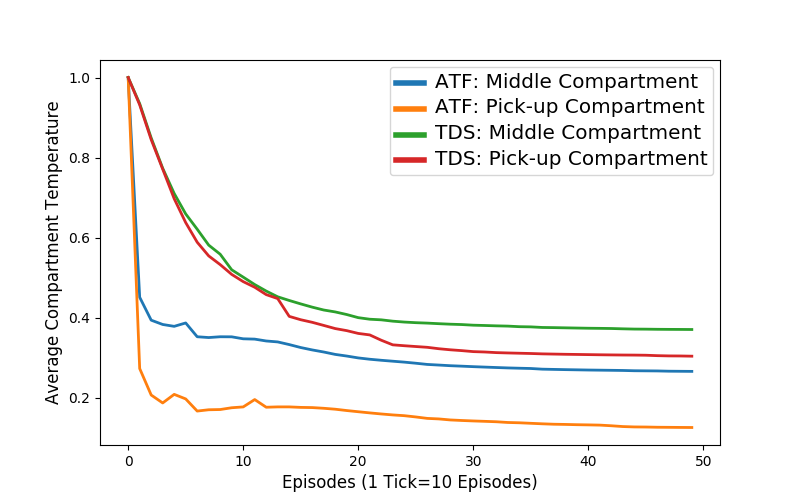} 
\caption{Average temperature per compartment}
\label{Deterministic_CMOTP_Results:LDQN_Vs_HDQN:AvgTemperaturePerCompartment}
\end{figure}

Lenient agents implemented with ATF struggle significantly within the narrow-passage CMOTP, as evident from the results listed in Table \ref{fig:originalCMOTP:Results} (right). We find that the average temperature values cool off rapidly over the first 100 episodes within the \textit{Pickup} and \textit{Middle} compartments, as illustrated in Figure {\ref{Deterministic_CMOTP_Results:LDQN_Vs_HDQN:AvgTemperaturePerCompartment}}. Meanwhile agents using TDS manage to maintain sufficient leniency over the first 1000 episodes to allow rewards to propagate backwards from the terminal state. We conducted ATF experiments with a range of values for the fold-in constant $\upsilon$ (0.2, 0.4 and 0.8), but always witnessed the same outcome. Slowing down the temperature decay would help agents using ATF remain lenient for longer, with the side-effects of an overoptimistic disposition in stochastic environments, and an increase in the number of steps required for convergence if the temperatures are tied to the action selection policy. Using TDS meanwhile allows agents to maintain sufficient leniency around difficult sub-tasks within the environment while being able to decay later transitions at a faster rate. As a result agents using TDS can learn the average rewards for state transitions close to the terminal state while remaining optimistic for updates to earlier transitions.



The success of HDQN agents within the narrow-passage CMOTP depends on the value chosen for $\beta$. Agents with $\beta > 0.5$ struggle to coordinate, as we observed over a large range of $\beta$ values, exemplars of which are given in Table \ref{fig:originalCMOTP:Results}.  The only agents that converge upon a near optimal joint-policy are those using LDQN-TDS and HDQN ($\beta=0.5$). We performed a Kolmogorov-Smirnov test with a null hypothesis that there is no significant difference between the performance metrics for agents using LDQN-TDS and HDQN ($\beta=0.5$). We fail to reject the null hypothesis for average steps per episode and percentage of coordinated steps for the final 100 episodes. However, HDQN ($\beta=0.5$) averaged significantly less steps per run while maintaining less overhead, replicating previous observations regarding the strengths of hysteretic agents within deterministic environments.


\section{Stochastic CMOTP Results}  \label{sec:stochastic_reward_cmotp}

In the stochastic setting we are interested in the percentage of runs for each algorithm that converge upon the optimal joint policy, which is for the agents to deliver the goods to dropzone 1, yielding a reward of 0.8, as opposed to dropzone 2 which only returns an average reward of 0.76 (see Section \ref{section:evaluation:cmotp}). We conducted 40 runs of 5000 episodes for each algorithm. 

As discussed in Section \ref{Deterministic_CMOTP_Results}, HDQN agents using $\beta > 0.7$ frequently fail to coordinate in the deterministic CMOTP. Therefore, setting $\beta=0.7$ is the most likely candidate to succeed at solving the stochastic reward CMOTP for standard HDQN architectures. However, agents using HDQN ($\beta=0.7$) only converged towards the optimal policy on 42.5\% of runs. The \textit{scheduled}-HDQN performed significantly better achieving a 77.5\% optimal policy rate. Furthermore the SHDQN performs well when an additional funnel-like narrow-passage is inserted close to the dropzones, with 93\% success rate. The drop in performance upon removing the funnel suggests that the agents are led astray by the optimism applied to earlier transitions within each episode, presumably around the pickup area where a crucial decision is made regarding the direction in which the goods should be transported. 

\begin{figure}[ht]
\centering
    \begin{subfigure}[b]{0.22\textwidth}
        \centering
    \includegraphics[width=\textwidth]{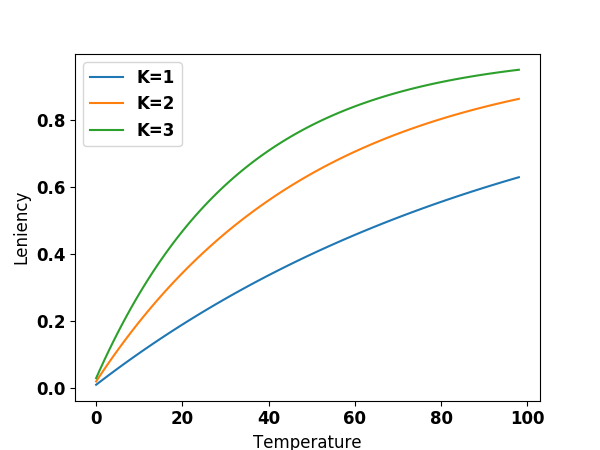} 
    \caption{Leniency Schedules}
    \label{fig:Stochastic_CMOTP_Results:Temperature_Based_Exploration:leniency_schedules}
    \end{subfigure}
    \begin{subfigure}[b]{0.22\textwidth}
        \centering
        \includegraphics[width=\textwidth]{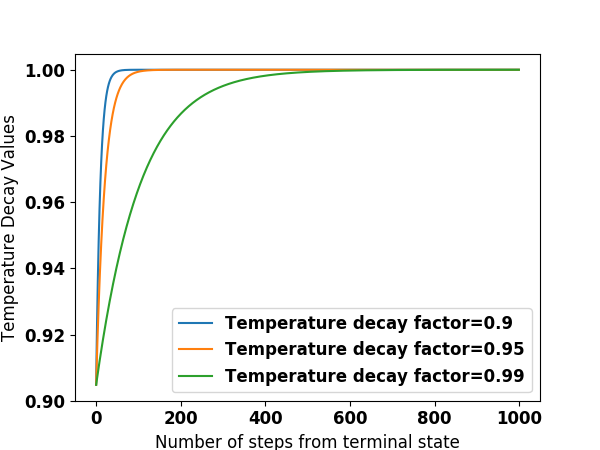} 
        \caption{TDS}
        \label{fig:Stochastic_CMOTP_Results:Temperature_Based_Exploration:TDSs}
    \end{subfigure}     
\caption{TMC and TDS schedules used during analysis.}
\label{fig:tmc_and_tds_schedules_for_analysis}
\end{figure}

LDQN using $\epsilon-Greedy$ exploration performed similar to SHDQN, converging towards the optimal joint policy on 75\% of runs. Meanwhile LDQNs using $\overline{T}(s_t)$-Greedy exploration achieved the highest percentages of optimal joint-policies, with agents converging on 100\% of runs for the following configuration: $K=3.0$, $d=0.9$, $\xi=0.25$ and $\mu=0.9995$, which will be discussed in more detail below. However the percentage of successful runs is related to the choice of hyperparameters. We therefore include an analysis of three critical hyperparameters:

\begin{itemize}
\item {The temperature Modification Coefficient $K$, that determines the speed at which agents transition from optimist to average reward learner (sub-figure \ref{fig:Stochastic_CMOTP_Results:Temperature_Based_Exploration:leniency_schedules}). \textbf{Values: 1, 2 and 3}}
\item {The TDS decay-rate $d$ which controls the rate at which temperatures are decayed $n$-steps prior to the terminal state (sub-figure \ref{fig:Stochastic_CMOTP_Results:Temperature_Based_Exploration:TDSs}). \textbf{Values: 0.9, 0.95 and 0.99}}
\item{$\overline{T}(s_t)$-Greedy exploration exponent $\xi$, controlling the agent's transition from explorer to exploiter, with lower values for $\xi$ encouraging exploration. \textbf{Values: 0.25, 0.5 and 1.0}}
\end{itemize}

\begin{figure}[tb]
\centering
    \begin{subfigure}[b]{0.15\textwidth}
        \centering
    \includegraphics[height=2.3cm]{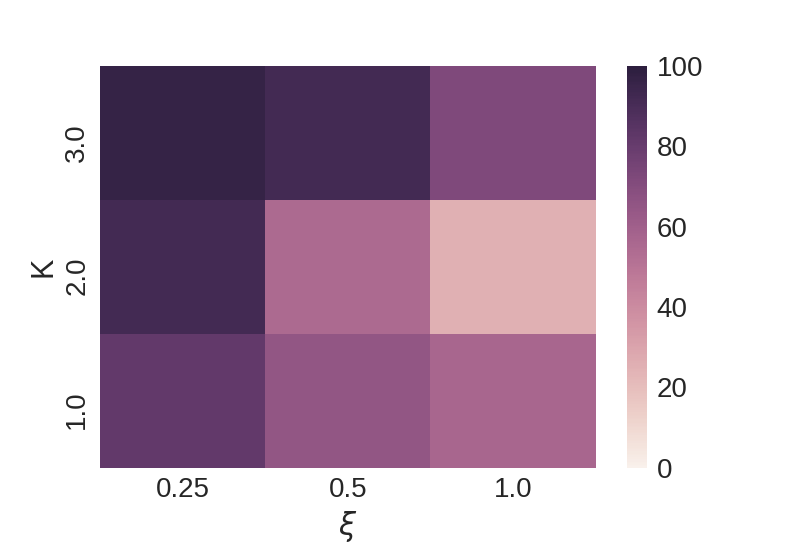} 
    \caption{$d=0.9$}
    \label{fig:Stochastic_CMOTP_Results:LDQN_Hyperparameter_Analysis:TDF_09}
    \end{subfigure}
    \begin{subfigure}[b]{0.15\textwidth}
        \centering
    \includegraphics[height=2.3cm]{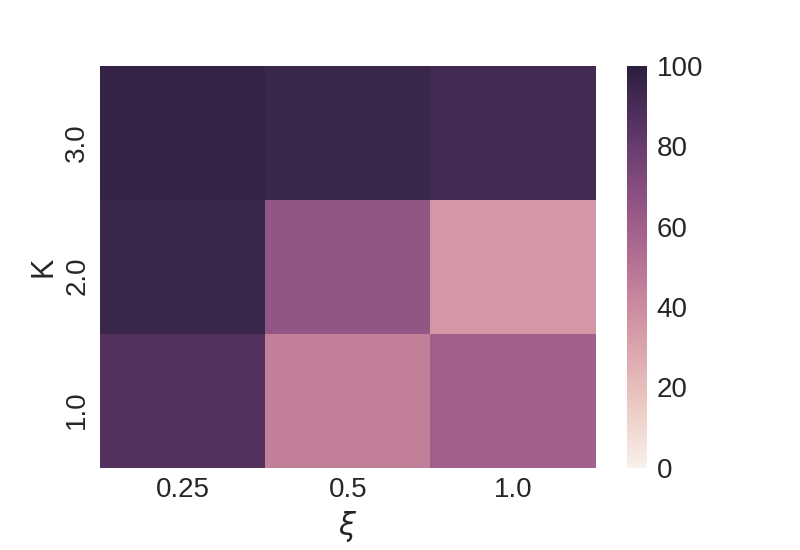} 
    \caption{$d=0.95$}
    \label{fig:Stochastic_CMOTP_Results:LDQN_Hyperparameter_Analysis:TDF_095}
    \end{subfigure}
    \begin{subfigure}[b]{0.14\textwidth}
        \centering
        \includegraphics[height=2.3cm]{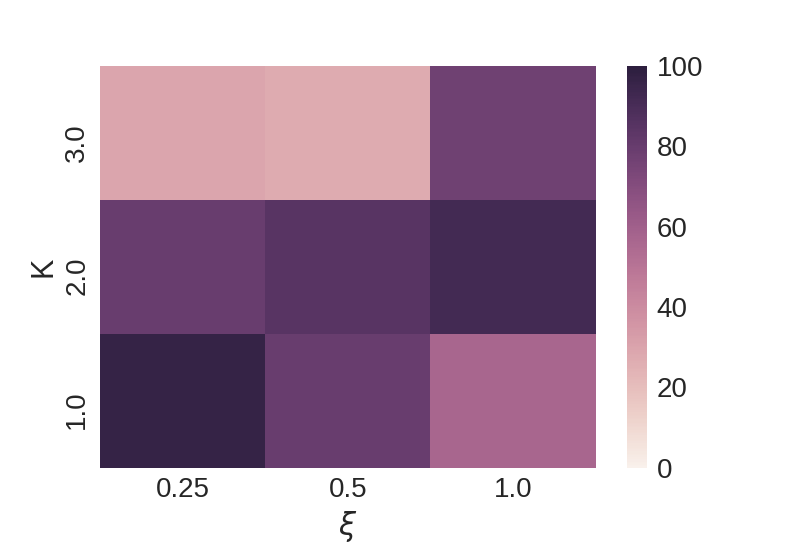} 
        \caption{$d=0.99$}
        \label{fig:Stochastic_CMOTP_Results:LDQN_Hyperparameter_Analysis:TDF_099}
    \end{subfigure}     
\caption{Analysis of the LDQN hyperparameters. The heat-maps show the percentage of runs that converged to the optimal joint-policy (darker is better).}
\label{fig:Stochastic_CMOTP_Results:LDQN_Hyperparameter_Analysis}
\end{figure}

We conducted 40 simulation runs for each combination of the three variables. To determine how well agents using LDQN can cope with stochastic state transitions we added a slippery surface where each action results in a random transition with 10\% probability \footnote{Comparable results were obtained during preliminary trials without a slippery surface.}. The highest performing agents used a steep temperature decay schedule that maintains high temperatures for early transitions ($d=0.9$ or $d=0.95$) with temperature modification coefficients that slow down the transition from optimist to average reward learner ($K=2$ or $k=3$), and exploration exponents that delay the transition from explorer to exploiter ($\xi=0.25$ or $\xi=0.5$). This is illustrated in the heat-maps in Figure \ref{fig:Stochastic_CMOTP_Results:LDQN_Hyperparameter_Analysis}. When using a TDS with a more gradual incline ($d=0.99$) the temperature values from earlier state transitions decay at a similar rate to those near terminal states. In this setting choosing larger values for $K$ increases the likelihood of the agents converging upon a sub-optimal policy prior to having established the average rewards available in later states, as evident from the results plotted in sub-figure {\ref{fig:Stochastic_CMOTP_Results:LDQN_Hyperparameter_Analysis:TDF_099}}.  Even when setting the exploration exponent $\xi$ to $0.25$ the agents prematurely transition to exploiter while holding an overoptimistic disposition towards follow-on states. Interestingly when $K<3$ agents often converge towards the optimal joint-policy despite setting $d=0.99$. However, the highest percentages of optimal runs (97.5\%) were achieved through combining a steep TDS ($d=0.9$ or $d=0.95$) with the slow transition to average reward learner ($k=3$) and exploiter ($\xi=0.25$). Meanwhile the lowest percentages for all TDSs resulted from insufficient leniency ($K=1$) and exploration ($\xi=1.0$). 

Using one of the best-performing configuration ($K=3.0$, $d=0.9$ and $\xi=0.25$) we conducted further trials analyzing the agents' sensitivity to the maximum temperature decay coefficient $\mu$. We conducted an additional set of 40 runs where $\mu$ was increased from 0.999 to 0.9995. Combining $\overline{T}(S)$-Greedy with the slow decaying $\mu=0.9995$ results in the agents spending more time exploring the environment at the cost of requiring longer to converge, resulting in an additional 1'674'106 steps on average per run. However, the agents delivered the best performance, converging towards the optimal policy on 100\% runs conducted.

\paragraph{\bf Continuous State Space Analysis} \label{Stochastic_CMOTP_Results:Continous_State_Space_Analysis}

Finally we show that semantically similar state-action pairs can be mapped to temperature values using SimHash in conjunction with an autoencoder. We conducted experiments in a noisy version of the stochastic CMTOP, where at each time step every pixel value is multiplied by a unique coefficient drawn from a Gaussian distribution $X \sim \mathcal{N}(1.0,0.01)$. A non-sparse tensor is used to represent the environment, with background cells set to 1.0 prior to noise being applied.

\begin{figure}[ht]
\centering
\includegraphics[width=0.4\textwidth]{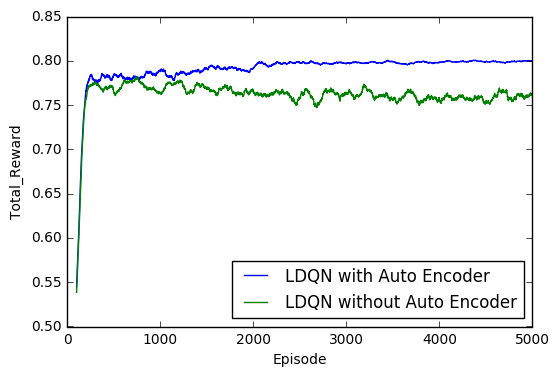} 
\caption{Noisy Stochastic CMOTP Average Reward}
\label{fig:Stochastic_CMOTP_Results:Continous_State_Space_Analysis:noisy_cmotp_results}
\end{figure}

Agents using LDQNs with xxhash converged towards the sub-optimal joint policy after the addition of noise as illustrated in Figure {\ref{fig:Stochastic_CMOTP_Results:Continous_State_Space_Analysis:noisy_cmotp_results}}, with the temperature values decaying uniformly in tune with $\nu$. LDQN-TDS agents using an autoencoder meanwhile converged towards the optimal policy on 97.5\% of runs. It is worth pointing out that the autoencoder introduces a new set of hyperparameters that require consideration, including the size $D$ of the dense layer at the centre of the autoencoder and the dimensions $K$ of the hash-key, raising questions regarding the influence of the granularity on the convergence. We leave this for future work.

\section{Discussion \& Conclusion}

Our work demonstrates that leniency can help MA-DRL agents solve a challenging fully cooperative CMOTP using high-dimensional and noisy images as observations. Having successfully merged leniency with a Double-DQN architecture raises the question regarding how well our LDQN will work with other state of the art components. We have recently conducted preliminary stochastic reward CMOTP trials with agents using LDQN with a \emph{Prioritized Experience Replay Memory} \cite{TomSchaul}. Interestingly the agents consistently converged towards the sub-optimal joint policy. We plan to investigate this further in future work. In addition our research raises the question how well our extensions would perform in environments where agents receive stochastic rewards throughout the episode. To answer this question we plan to test our LDQN within a hunter prey scenario where each episode runs for a fixed number of time-steps, with the prey being re-inserted at a random position each time it is caught \cite{matignon2012independent}. Furthermore we plan to investigate how our LDQN responds to environments with more than two agent by conducting CMOTP and hunter-prey scenarios with four agents.

To summarize our contributions:\\
\textbf{1)}  In this work we have shown that leniency can be applied to MA-DRL, enabling agents to converge upon optimal joint policies within fully-cooperative environments that require implicit coordination strategies and yield stochastic rewards.\\
\textbf{2)}  We find that LDQNs significantly outperform standard and scheduled-HDQNs within environments that yield stochastic rewards, replicating findings from tabular settings \cite{JMLR:v17:15-417}.\\
\textbf{3)}  We introduced two extensions to leniency, including a retroactive temperature decay schedule that prevents the premature decay of temperatures for state-action pairs and a $\overline{T}(s_t)$-Greedy exploration strategy that encourages agents to remain exploratory in states with a high average temperature value. The extensions can in theory also be used by lenient agents within non-deep settings.\\
\textbf{4)}  Our LDQN hyperparameter analysis revealed that the highest performing agents within stochastic reward domains use a steep temperature decay schedule that maintains high temperatures for early transitions combined with a temperature modification coefficient that slows down the transition from optimist to average reward learner, and an exploration exponent that delays the transition from explorer to exploiter.\\
\textbf{5)}  We demonstrate that the CMOTP {\cite{bucsoniu2010multi}} can be used as a benchmarking environment for MA-DRL, requiring reinforcement learning agents to learn fully-cooperative joint-policies from processing high dimensional and noisy image observations.\\ 
\textbf{6)}  Finally, we introduce two extensions to the CMOTP. First we include narrow passages, allowing us to test lenient agents' ability to prevent the premature decay of temperature values. Our second extension introduces two dropzones that yield stochastic rewards, testing the agents' ability to converge towards an optimal joint-policy while receiving misleading rewards.


\section{ Acknowledgments}
We thank the HAL Allergy Group for partially funding the PhD of Gregory Palmer and gratefully acknowledge the support of NVIDIA Corporation with the donation of the Titan X Pascal GPU that enabled this research.


\clearpage

\end{document}